\documentclass[a4paper,11pt]{article}
\pdfoutput=1 

\usepackage{jcappub} 

\usepackage[T1]{fontenc} 

\title{\boldmath Modeling gamma ray production from proton-proton interactions in high-energy astrophysical environments}

\author[a,1]{Dimitra Atri,\note{Corresponding author.}}
\author[b]{B. Hariharan}

\affiliation[a]{Particle Astrophysics Group, Blue Marble Space Institute of Science\\Seattle, WA 98145-1561, USA}
\affiliation[b]{Tata Institute of Fundamental Research, Department of High Energy Physics\\Colaba, Mumbai 400 005, India}

\emailAdd{dimitra@bmsis.org}
\emailAdd{89hariharan@gmail.com}

\abstract{Gamma rays are the best probes to study high-energy particle interactions occurring in astrophysical environments. Space based instruments such as Fermi Large Area Telescope (Fermi LAT) and ground based experiments such as VERITAS, H.E.S.S. and MAGIC have provided us with valuable data on various production mechanisms of gamma rays within our Galaxy and beyond. Depending on astronomical conditions, gamma rays can be produced either by hadronic or leptonic interactions. In this paper, we probe the production of gamma rays by the hadronic channel where gamma rays are primarily produced by the decay of secondary neutral pions and $\eta$ mesons from proton-proton interactions in a wide energy range. We use state of the art high-energy hadronic interaction models, calibrated with the new LHC results and widely used in ground based ultra-high energy air shower experiments. We also compare SIBYLL 2.1, QGSJET-II-04 and EPOS LHC models and provide lookup tables which can be used by researchers to model gamma ray production from the hadronic channel and ultimately extract the underlying proton spectrum from gamma ray observations.}

\begin{document}
\maketitle
\flushbottom

\section{Introduction}
\label{sec:intro}

Gamma ray astronomy provides insights into the physics of high-energy particle interactions occurring in space. Unlike ordinary photons, gamma rays can only be produced by the acceleration of charged particles in high-energy environment. Such conditions usually occur in astrophysical shocks where charged particles are accelerated to high-energies in a sudden energy burst, such as supernovae, gamma ray bursts etc. Accelerated particles can gain energies of the order of multi TeV and higher, depending on astrophysical conditions. These particles often interact with the interstellar medium or dense molecular clouds where particle collisions occur. This result in the production of secondary particles, some of which can be detected on Earth. Among the variety of secondary particles produced, neutral pions and $\eta$ mesons quickly decay to gamma rays and form the primary source of gamma ray observations. Equations 1.1 to 1.5 shows the various decay processes in detail. 
\begin{center}
\begin{equation}
$$pp $\rightarrow$ $\pi^{0}$ $\rightarrow$ 2$\gamma$ $$
\end{equation}
\begin{equation}
$$ $\eta$ $\rightarrow$ 2$\gamma$ (39.4\%)  $$        
\end{equation}
\begin{equation}
$$ $\eta$ $\rightarrow$ 3$\pi^{0}$ $\rightarrow$ 6$\gamma$ (32.5\%)$$
\end{equation}
\begin{equation}
$$ $\eta$ $\rightarrow$ $\pi^{+}$$\pi^{-}$$\pi^{0}$ (22.6\%)$$
\end{equation}
\begin{equation}
$$ $\eta$ $\rightarrow$ $\pi^{+}$$\pi^{-}$$\gamma$ (5\%)$$
\end{equation}

\end{center}
These gamma rays, depending on the energy, can be observed with both space and ground based instruments. The real challenge is to subsequently extract the particle spectra from gamma ray observations. To achieve this, a robust model of hadronic interactions is necessary, which can accurately produce gamma ray spectra. The simple analytical approach used in the earlier work \cite{kelner} was based on our knowledge of pre-LHC hadronic interactions and therefore, is inadequate to accurately reproduce particle distributions. In this paper, we model the proton-proton interactions and the resulting energy distribution of photons using such a model. We present the results in the form of lookup tables using which, the observed gamma ray spectrum can be reproduced after using appropriate spectral index, primary energy and normalization. We also compare hadronic interaction models by using standard particle interaction models to generate events for this purpose. 

The proton-proton interactions have been measured precisely up to $\sim$ PeV energies with accelerator experiments. Good amount of data is available to calibrate different hadronic interaction event generators, often referred to as hadronic interaction models. Hadronic interaction models have been developed and upgraded over the years to accurately reproduce particle production observed in hadronic interactions, with accelerators as well as with air shower experiments. Recently, a large quantity of useful data from LHC was made available to test the consistency of these models at higher energies. It was found that models dedicated to produce high energy physics data such as PYTHIA \cite{pythia} were not able to accurately reproduce strange particle production and distributions of lower transverse momenta particles \cite{epos-lhc}. Therefore, we use the EPOS LHC model \cite{epos-lhc}, which is able reproduce the LHC results to desired accuracy and compare it with other interaction models used in ultra-high energy cosmic ray experiments.
 
\begin{figure}[tbp]
\centering 
\includegraphics[width=0.85\textwidth]{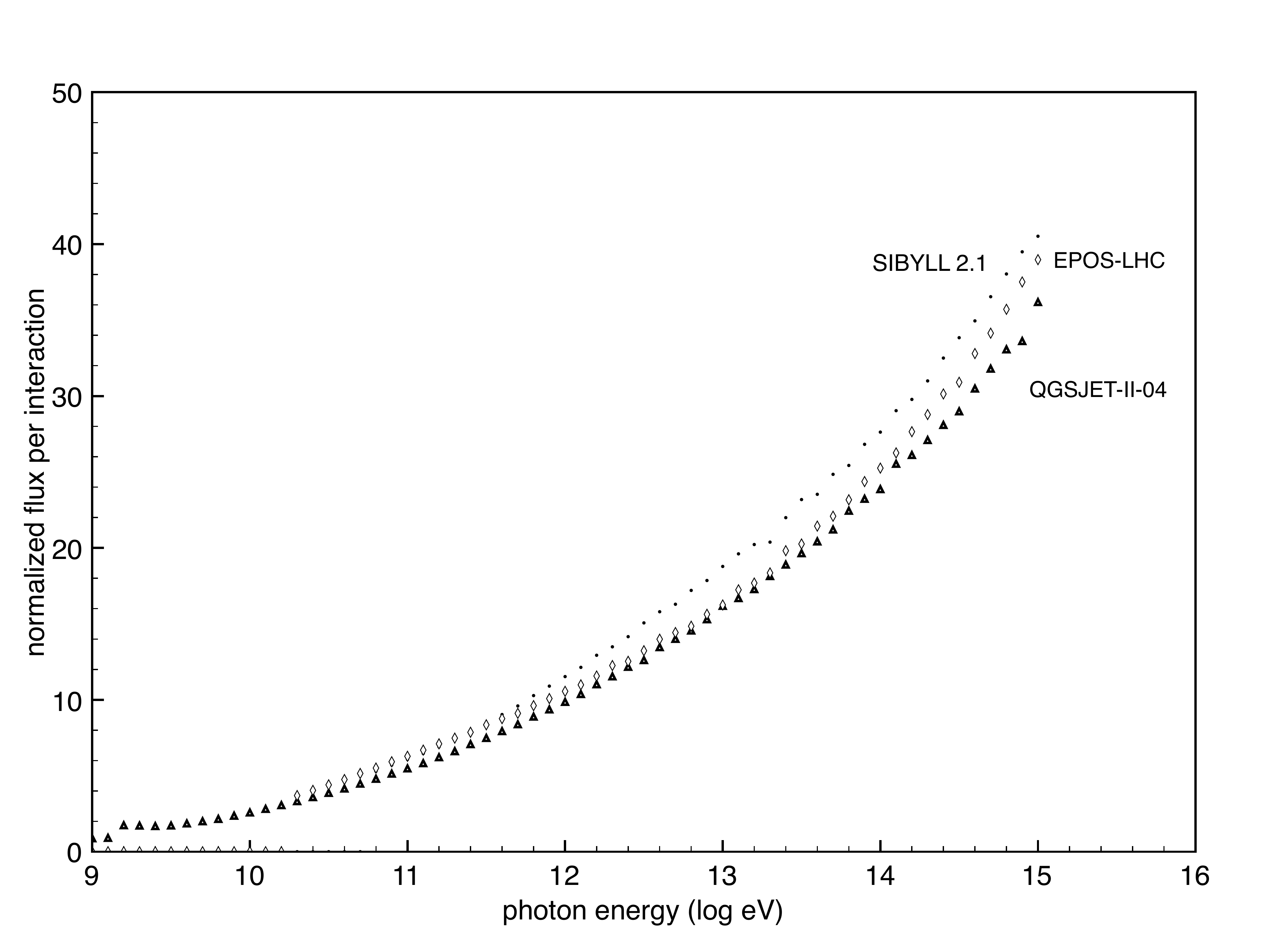}
\caption{\label{fig:i} Normalized total number of photons from proton-proton collisions from SIBYLL 2.1, QGSJET-II-04 and EPOS LHC model as a function of energy.}
\end{figure}

\begin{figure}[tbp]
\centering 
\includegraphics[width=0.85\textwidth]{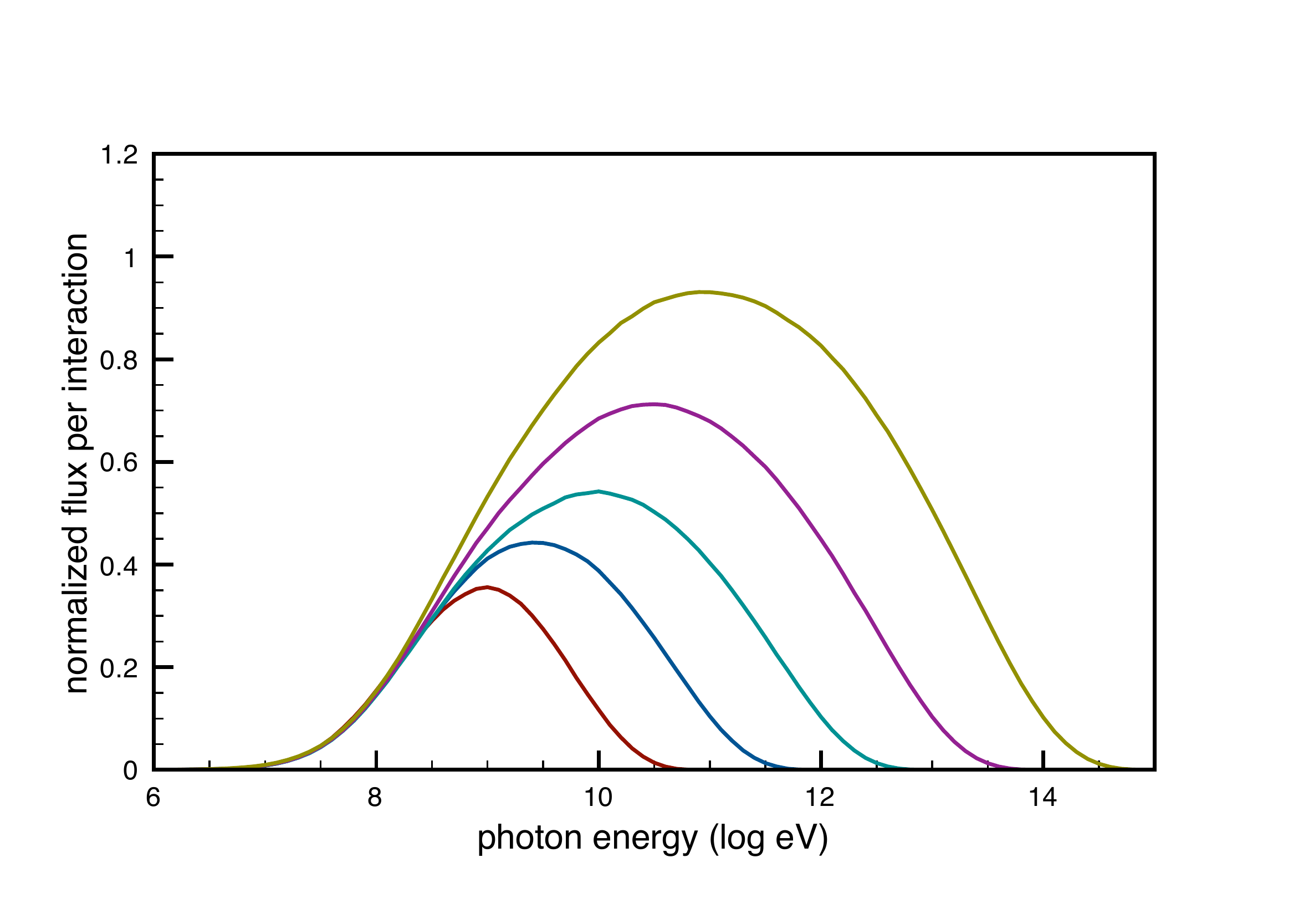}
\caption{\label{fig:i} Gamma ray spectra from the EPOS LHC model for 100 GeV (lowest curve), 1 TeV, 10 TeV, 100 TeV and 1 PeV (largest curve) primary interactions.}
\end{figure}
\section{Method}
We generate proton-proton interaction events by using the latest versions of various Monte Carlo event generators. The Cosmic Ray Monte Carlo (CRMC 1.0) \cite{crmc} package was used as a common interface to interact with all the event generators. Most of the event generators are in FORTRAN and each has a separate interface to carry out computations. Also each event generator has their own I/O and particle code conversion. CRMC provides common C++ interface to these FORTRAN sources. Output can be recorded either in HepMC or ROOT format which is commonly used in high energy physics. It uses a standard common particle code identification known as PDG(Particle Data Group) IDs. 

In this work, we have used SIBYLL 2.1 \cite{sibyll}, QGSJET-II-04 \cite{qgs} and EPOS-LHC \cite{epos-lhc} Monte Carlo event generators. A large number of events are generated for each energy so that numerical errors are reduced to the least. The EPOS LHC model is a modified and upgraded version of the EPOS 1.99 model \cite{epos1}, and has been calibrated with the latest results from LHC \cite{epos-lhc}. All three event generators have been tested against all major air shower experiments, and EPOS LHC has been recently calibrated with the LHC results. 

The tables are constructed with primary energies starting with 1 GeV up to 1 PeV with the logarithmic bin width of 0.1. $10^{6}$ events are generated for each energy with each model and the normalised gamma ray flux is recorded and displayed in the form of look-up tables. The photon energies are recorded in logarithmic bins with the bin size of 0.1. All units displayed in the tables are in eV and all logarithmic values are with respect to base 10. Finally the table is normalised to per primary.  One can use these tables to get the gamma ray spectrum after an assumed proton spectrum instead of running different event generators in different ways.

\begin{figure}[tbp]
\centering 
\includegraphics[width=0.85\textwidth]{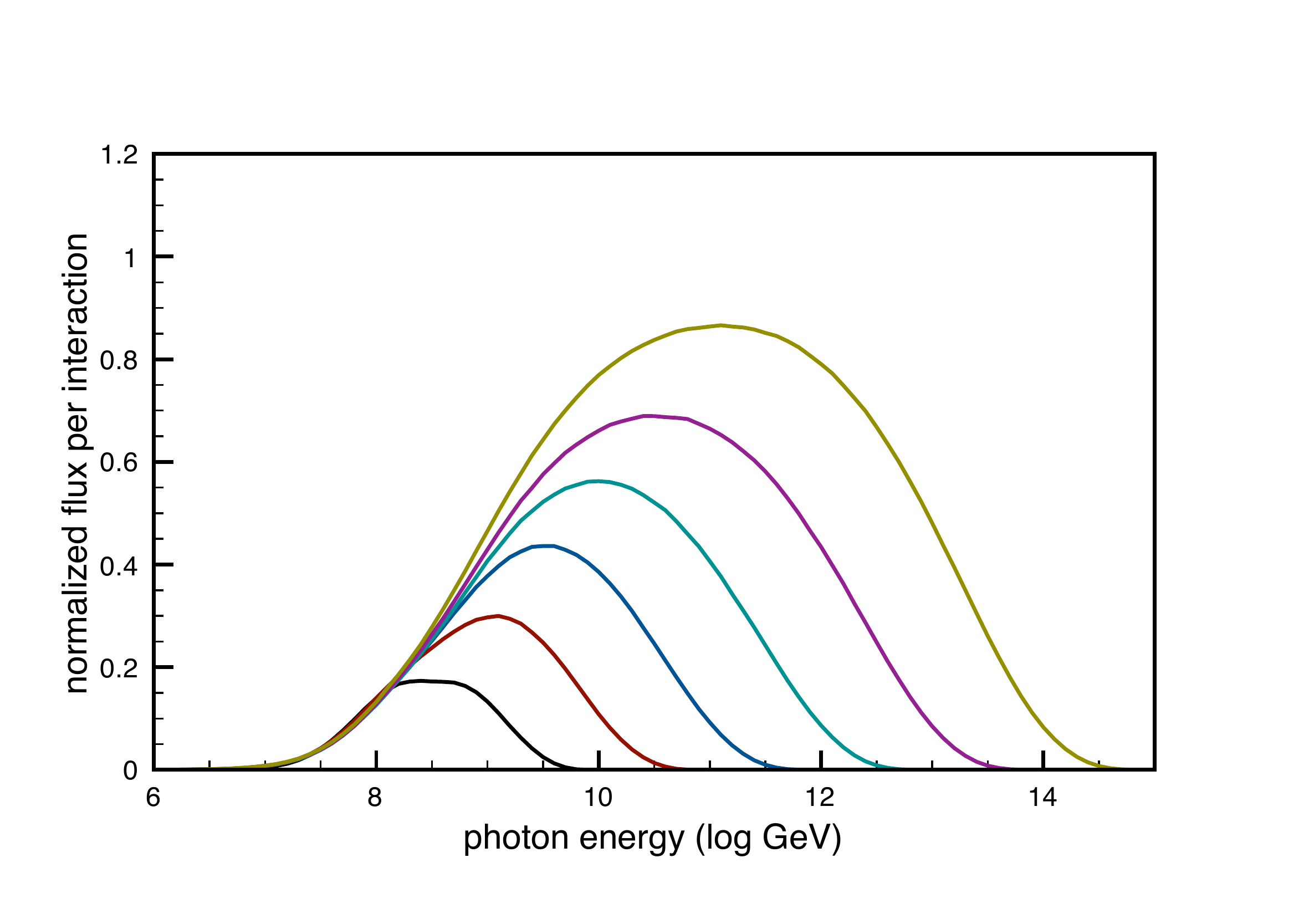}
\caption{\label{fig:i} Gamma ray spectra from the QGSJET-II-04 model for 10 GeV (lowest curve), 100 GeV, 1 TeV, 10 TeV, 100 TeV and 1 PeV (largest curve) primary interactions.}
\end{figure}

\begin{figure}[tbp]
\centering 
\includegraphics[width=0.85\textwidth]{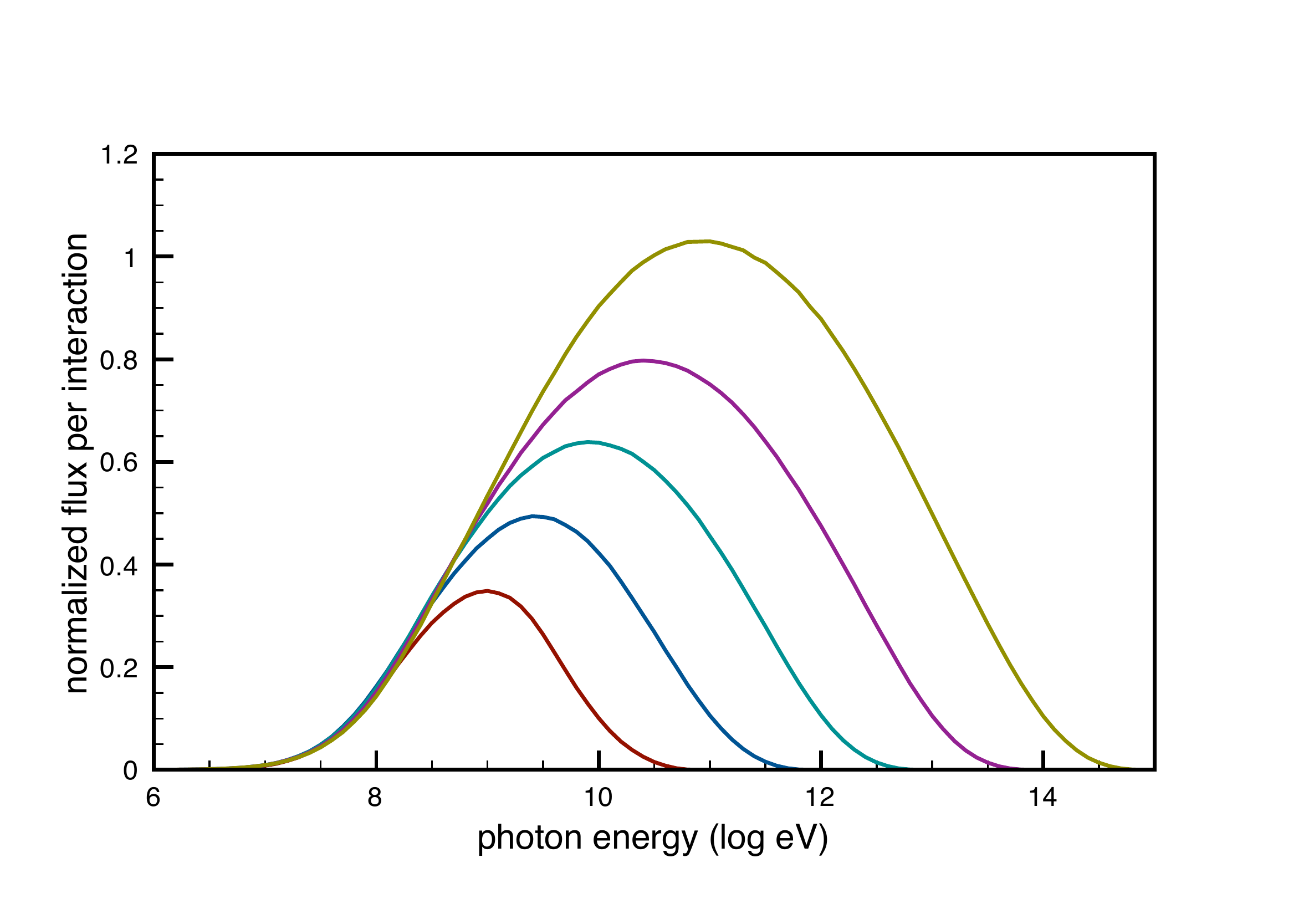}
\caption{\label{fig:i} Gamma ray spectra from the SIBYLL 2.1 model for 100 GeV (lowest curve), 1 TeV, 10 TeV, 100 TeV and 1 PeV (largest curve) primary interactions.}
\end{figure}

\section{Results}
The energy spectrum for each energy is obtained from each model and is recorded in the form of lookup tables giving the normalised flux as a function of photon energy. As expected, more number of photons are produced per interaction as the primary energy is increased. Also, the peak of the energy spectrum shifts to higher energy with an increase in the primary energy. These features can be seen in figures 2, 3 and 4. At first glance, all three models seem to give similar results, however upon close inspection, differences can be seen. 

\begin{figure}[tbp]
\centering 
\includegraphics[width=0.85\textwidth]{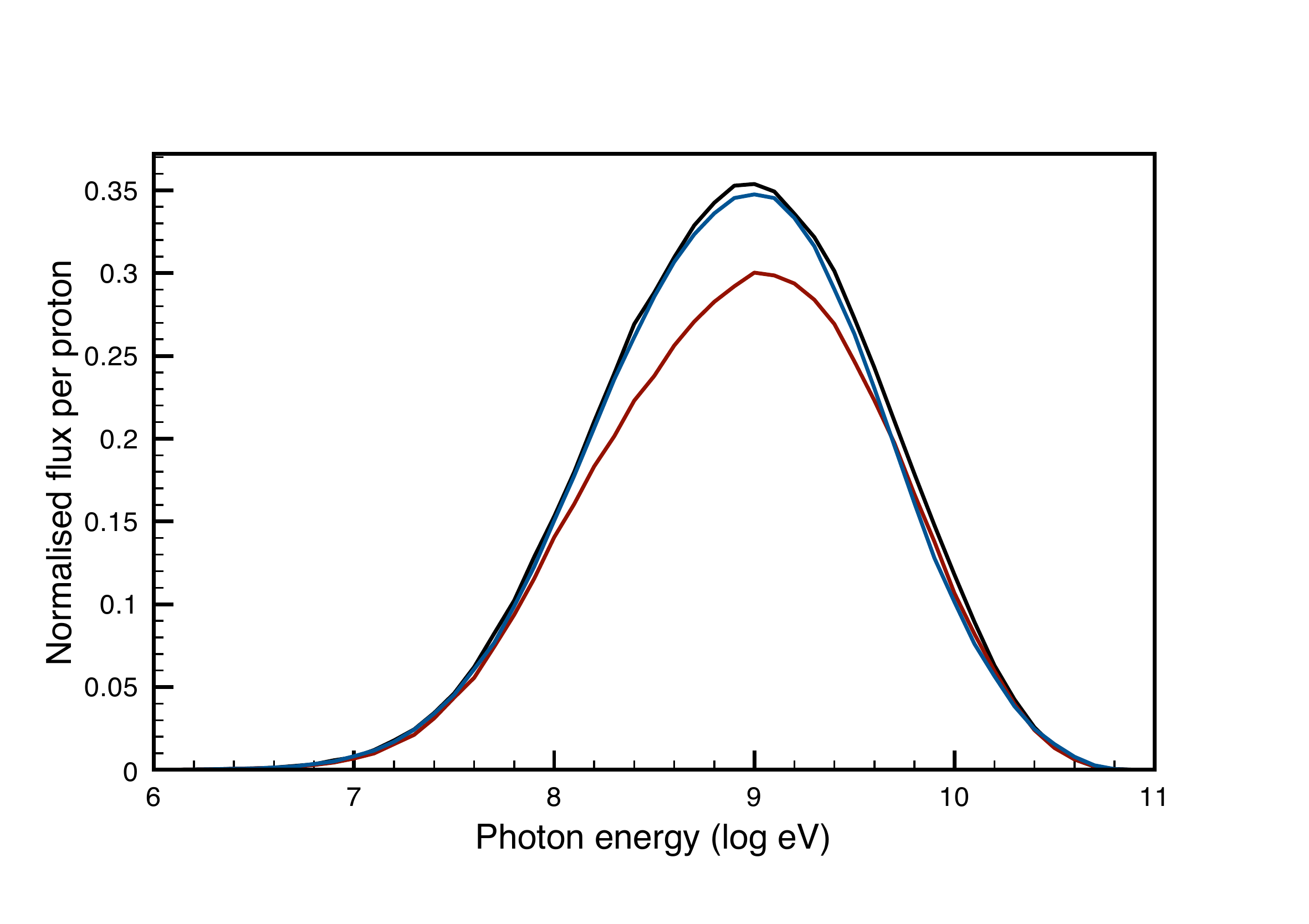}
\caption{\label{fig:i} Comparison of gamma ray spectra with different models at 100 GeV. SIBYLL 2.1 (blue), EPOS LHC (black) and QGSJET-II-04 (red).}
\end{figure}

\begin{figure}[tbp]
\centering 
\includegraphics[width=0.85\textwidth]{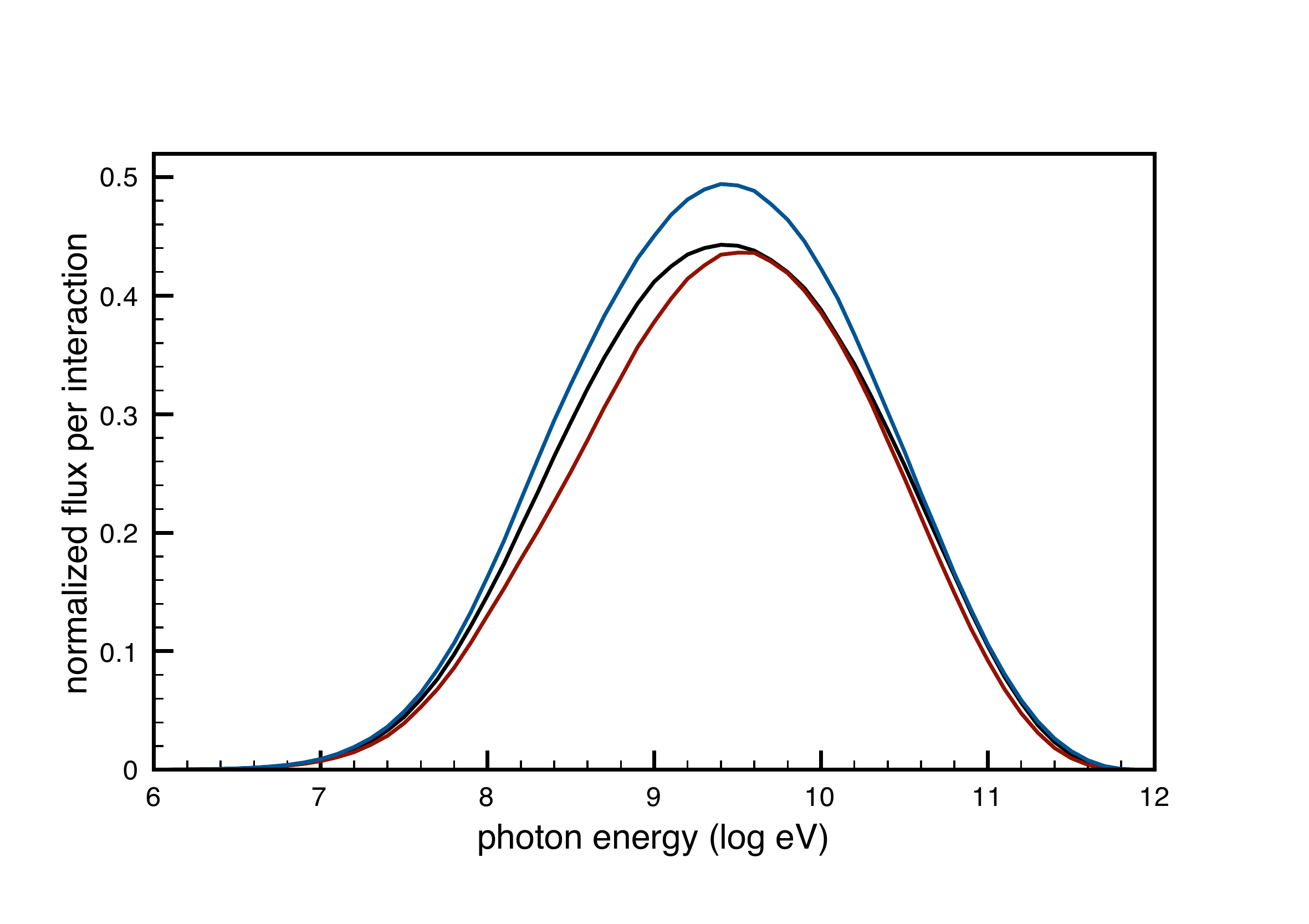}
\caption{\label{fig:i} Comparison of gamma ray spectra with different models at 1 TeV. SIBYLL 2.1 (blue), EPOS LHC (black) and QGSJET-II-04 (red).}
\end{figure}

\begin{figure}[tbp]
\centering 
\includegraphics[width=0.85\textwidth]{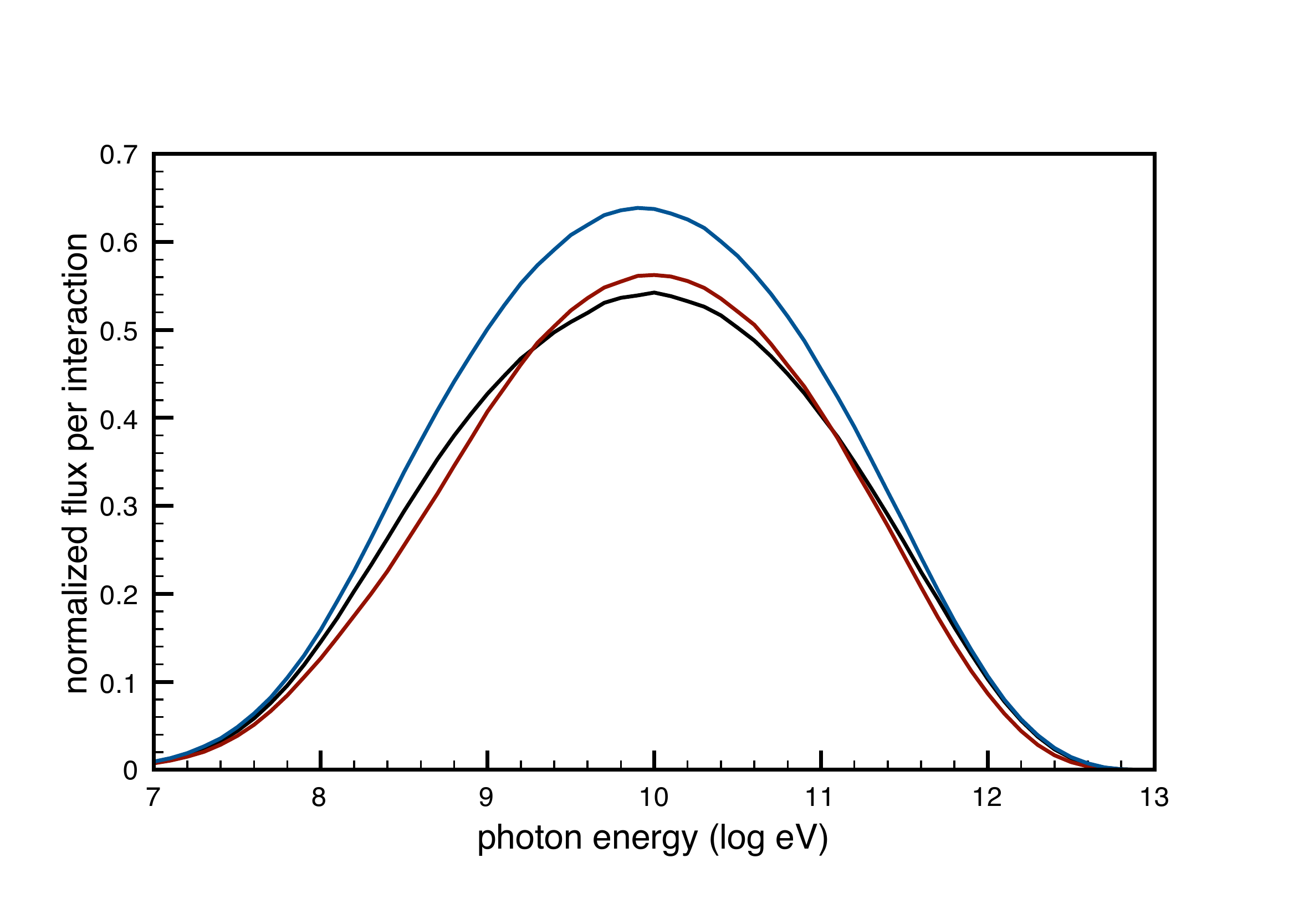}
\caption{\label{fig:i} Gamma ray spectra from the three different models at 10 TeV primary interaction. SIBYLL 2.1 (blue), EPOS LHC (black) and QGSJET-II-04 (red).}
\end{figure}

\begin{figure}[tbp]
\centering 
\includegraphics[width=0.85\textwidth]{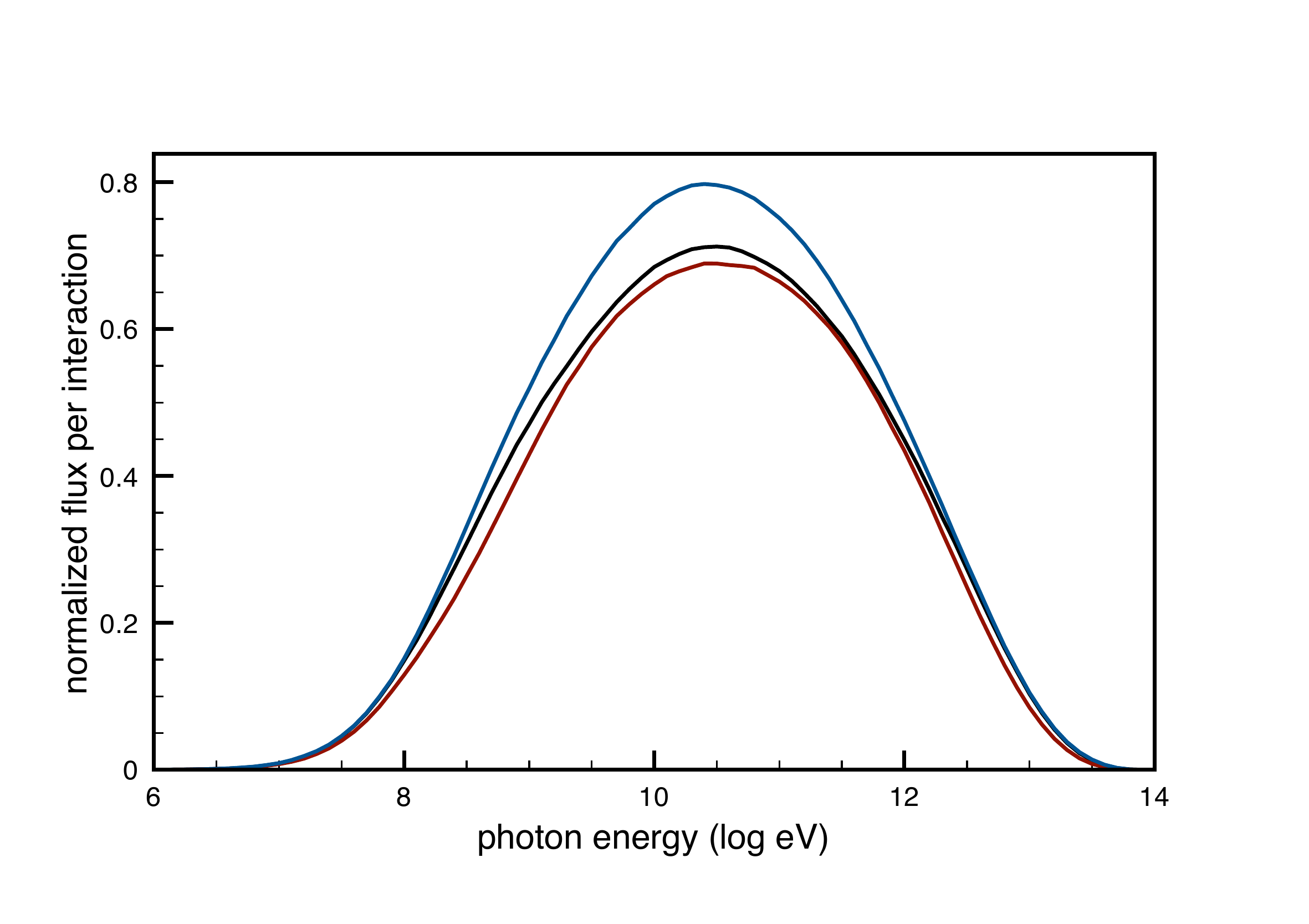}
\caption{\label{fig:i} Gamma ray spectra from the three different models at 100 TeV primary interaction. SIBYLL 2.1 (blue), EPOS LHC (black) and QGSJET-II-04 (red).}
\end{figure}

\begin{figure}[tbp]
\centering 
\includegraphics[width=0.85\textwidth]{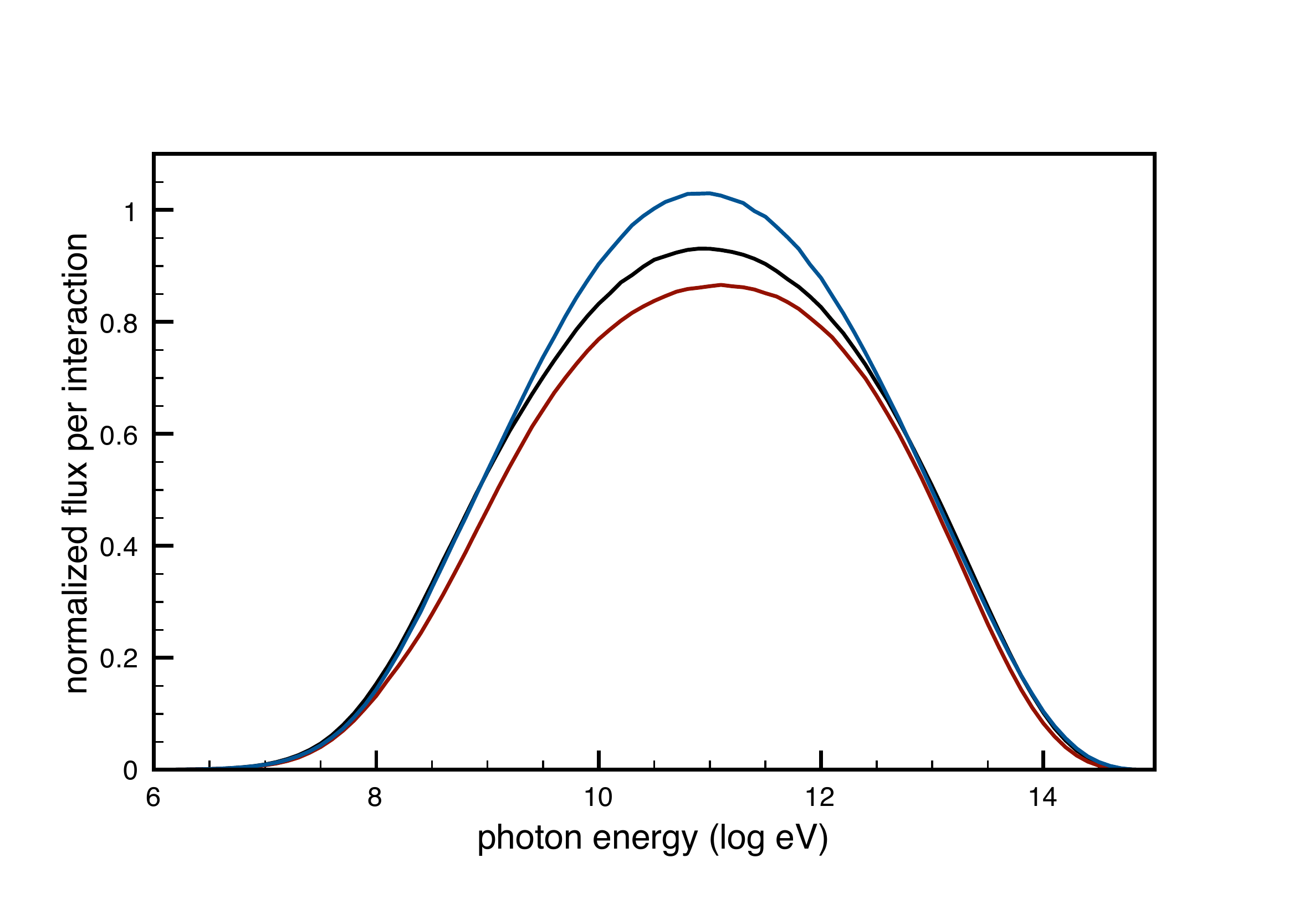}
\caption{\label{fig:i} Gamma ray spectra from the three different models at 1 PeV primary interaction. SIBYLL 2.1 (blue), EPOS LHC (black) and QGSJET-II-04 (red).}
\end{figure}

\begin{figure}[tbp]
\centering 
\includegraphics[width=0.85\textwidth]{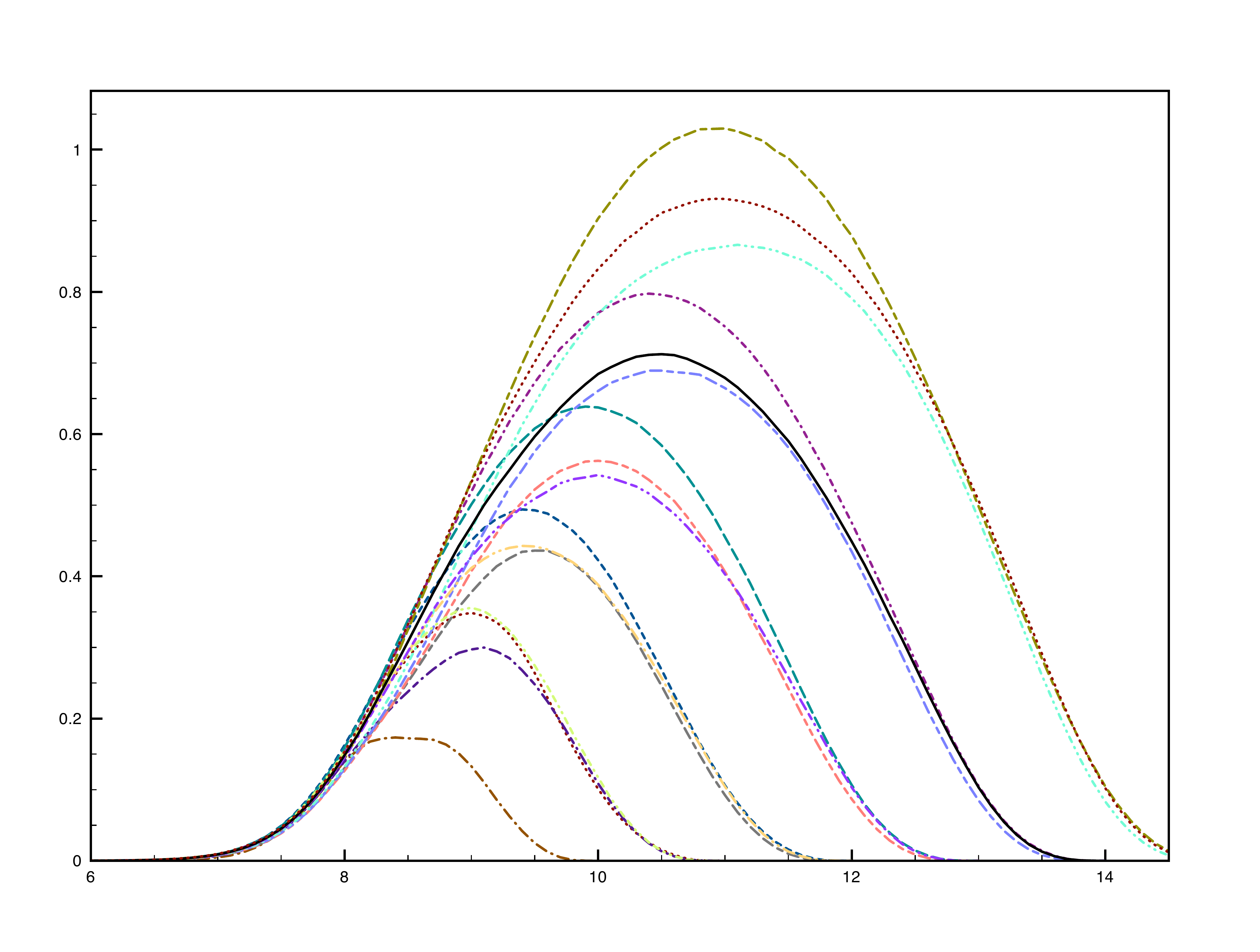}
\caption{\label{fig:i} Comparison of gamma ray spectra with different models between 10 GeV - 1 PeV. }
\end{figure}

At 1 PeV, 100 TeV, and 1 TeV, SIBYLL 2.1 gives the highest number of photons, followed by EPOS LHC and QGSJET-II-04. At 10 TeV, the highest flux is given by SIBYLL 2.1 as at higher energies and is followed by QGSJET-II-04 and EPOS LHC. However, at 100 GeV, EPOS LHC leads the flux followed by SIBYLL 2.1 and QGSJET-II-04. At 10 GeV and below, only the SIBYLL 2.1 model gave outputs as shown in figure 1. Next, we comment on the peak of the energy spectrum at particular energies. At 1 PeV, 100 TeV, 1 TeV and 100 GeV, QGSJET-II-04 shows the spectrum peaking at higher gamma ray energy, whereas the other two models have the peak shifted at similar but lower energies. At 10 TeV, both EPOS LHC and QGSJET-II-04 peak at the same energy and SIBYLL 2.1 peaks at lower energy. As a summary, a comparison of different models at different energies is shown in figure 10. 

One can choose a model to assume a source spectrum and use it to calculate number of protons per unit area per unit time. Since all values in lookup tables are normalized to a single proton-proton interaction, one  needs to multiply the number obtained from calculations to the numbers in the table to get the gamma ray spectrum in that energy range. 

\section{Discussion}
Gamma ray observations have become a powerful tool to probe the nature of underlying physics of high energy astrophysical processes. In this paper we have focused on the production of gamma rays through the hadronic channel. Particle acceleration can be measured indirectly using gamma rays by using models that compute gamma ray production resulting from hadronic interactions. However, it can be very time consuming to perform such calculations for each observation.  Here, we have presented lookup tables that can be used by researchers to compute the proton spectrum of astrophysical sources from gamma ray observations. We recommend the use of the lookup table produced with the EPOS LHC model since it has been calibrated with recent LHC results and thus, provides the best description of hadronic interactions at high energies. Tables will be made public on publication of the manuscript.

\acknowledgments
We thank Tanguy Pierog and Felix Reihn for their help in using interaction models.


\end{document}